# Scientists' Warning on Technology

Authors: Bill Tomlinson[1, 2*], Andrew W. Torrance[3,4], William J. Ripple[5]

[1] Department of Informatics, University of California, Irvine, Irvine, CA 92697 USA.

[2] School of Information Management, Victoria University of Wellington - Te Herenga Waka, Wellington, New Zealand.

[3] School of Law, University of Kansas, Lawrence, KS 66045 USA.

[4] Sloan School of Management, Massachusetts Institute of Technology, Cambridge, MA 02142 USA.

[5] Department of Forest Ecosystems and Society, Oregon State University, Corvallis, OR 97331 USA.

[*] Corresponding author. Email: wmt@uci.edu.

**In the past several years, scientists have issued a series of warnings about the threats of climate change and other forms of environmental disruption. Here, we provide a scientists' warning on how technology affects these issues. Technology simultaneously provides substantial benefits for humanity, and also profound costs. Current technological systems are exacerbating climate change and the wholesale conversion of the Earth's ecosystems. Adopting new technologies, such as clean energy technologies and artificial intelligence, may be necessary for addressing these crises. Such transformation is not without risks, but it may help set human civilizations on a path to a sustainable future.**

# Introduction

In 1992, the Union of Concerned Scientists issued a warning to humanity, writing: "great change in our stewardship of the [E]arth and the life on it is required, if vast human misery is to be avoided and our global home on this planet is not to be irretrievably mutilated",[1] In 2017, a team of scientists published a second warning,[2] updating the information provided by the 1992 notice; this document has been signed by more than 15,000 scholars across many different scientific fields.  Since that time, scholars around the world have begun to publish a series of "Scientists' Warning" articles, detailing perspectives from their particular disciplines to document the array of environmental issues currently threatening the Earth. To date, 39 articles have been published, and more than 50 more are currently in preparation. Collectively, these articles seek to offer a broad understanding of ways that humanity is undermining the foundation of its own existence.

Two key articles in the Scientists' Warning series–"Scientists' Warning on Population"[3] and "Scientists' Warning on Affluence"[4]–index into a set of concerns that rose to prominence several decades ago as part of the so-called I=PAT equation.[5,6] This equation describes environmental impact (I) as being a function of population (P), affluence (A), and technology (T). While technology is equally salient to these discussions as population and affluence, to date, there has not been an article providing a scientists' warning on technology.  This article seeks to fill that gap.

In this article, we have adapted the definition of technology originally suggested by Stanford scientist Brian Arthur[7], and define technology as "a quantifiable and duplicatable means to fulfill a human purpose." Technologies are typically developed by human civilizations to solve problems and produce benefits for people (e.g., the Haber-Bosch process to produce fertilizer for enhanced agricultural productivity, steam engines to manufacture and distribute consumer goods, automobiles to support large-scale mobility, etc.).  However, those very technologies create problems of their own (often called costs or externalities[8] in economics and related fields), via a variety of social and environmental impacts (water pollution, social inequality, climate change, etc.).  This dual nature of technology–that of both solving and creating problems–provides the underpinnings for the core argument of this article.

Here, we focus on two domains of technological innovation as exemplars of the range of forms technology can take: clean energy technologies[9,10] and artificial intelligence (AI).[11,12]  The benefits of clean energy technologies, such as electrification, regarding climate change and other environmental issues are well-established.[9,10]  The benefits of AI in these domains are promising but still emerging.[11,12]

While both of these technologies have great potential, they also entail various risks, from electronic waste and (temporary) job displacement for electrification technologies[9] to runaway energy use and existential threats from potential future superintelligences for AI.[13]

While technology has played a key role in both bringing about many of the global challenges humanity currently faces, and may also play a key role in mitigating them, we warn against placing too much faith in technology to solve such issues singlehandedly.[14] As a 2012 National Academies report titled Computing Research for Sustainability states: "Despite the profound technical challenges involved, sustainability is not, at its root, a technical problem, nor will merely technical solutions be sufficient. Instead, deep economic, political, and cultural adjustments will ultimately be required, along with a major, long-term commitment in each sphere to deploy the requisite technical solutions at scale. Nevertheless, technological advances and enablers have a clear role in supporting such change."[15] In that spirit, we offer three warnings relating to technology.

First, we warn that some current technologies are creating great harm through climate change and habitat loss, and that these technologies should be phased out as soon as possible. Accordingly, we encourage rapid and comprehensive decarbonization[10] and other technological shifts[11] to reduce the harms brought about by current technologies. This phase-out may be enabled by broad-scale adoption of electrification technologies, an embracing of the possibility of allowing AI to help coordinate and augment our abilities to address the global sustainability challenges facing humanity, and research and development on other technologies that may support these efforts. In addition, we also discuss the possibility of undesign[16], the act of intentionally extracting technology from a given context.

Second, we warn that future technologies are likely to produce harms of their own, only some of which we can predict in advance; while we encourage pursuing these new technologies with vigor as substitutions for current harmful technologies, we also encourage vigilance about the potential for future harm. Initiatives to pursue innovation should be intertwined with important threads of research about harm prevention. This article provides an array of perspectives emerging in technological fields about how to integrate rapid technological transformation with appropriate reduction of harmful externalities to achieve long-term societal goals. Nevertheless, even though new technologies will come with new harms, the harms of current technologies (e.g., climate change) are so severe that it is imperative to pursue new technologies that can supplant the old. Humanity should not allow the potential for future harm to prevent itself from pursuing critical technological change.

And finally, we warn that, although technology will almost certainly be a central part of a large-scale human response to environmental degradation, technology alone will not be enough; we need "deep economic, political, and cultural adjustments",[15] and new narratives that help people acclimate to new ways of life. There are efforts afoot to develop technologies that can support such efforts to develop and spread new narratives;[17] we encourage technological innovations that help envision sustainable futures and help people learn how to bring them into existence.

Technologies represent crucial tools that, along with economic, political, and cultural shifts, can help humanity address the looming issues threatening life on Earth. This article seeks to provide a theoretical framework for thinking about the various benefits and harms that arise from current and future technological systems, and how to integrate them with civilization-scale transitions to sustainability.

## The Role of Technology in I=PAT

The late 1960s and early 1970s witnessed vigorous debates among the public, government, and scholars about both the extent and causes of environmental degradation. The United States Congress and President Richard Nixon passed comprehensive federal statutes aimed at counteracting polluted air (the "Clean Air Act"[18]) and water (the "Clean Water Act"[19]), protecting biodiversity (the "Endangered Species Act"[20]), and correcting a lack of national environmental review (the "National Environmental Policy Act"[21]). Nevertheless, many scholars remained worried about impending ecological collapse. One prominent debate involved Barry Commoner, who argued that improvements in technology posed the most serious threat,[6] and Paul Ehrlich and John Holdren, who acknowledged the peril of continuous technological improvement, but also proposed that material consumption and human population all combined with technology to degrade the environment.[5]

To help clarify the terms of the debate, Ehrlich and Holdren adapted a simple mathematical formula that Commoner[6] had included in a footnote (p. 211-212) to yield an equation that attempted to illustrate the relationships among what they termed population, affluence, and technology. This equation[5] is as follows:

$$I = P * A * T$$

For Ehrlich and Holdren, I represented impact on the environment, P the human population, A affluence (that is, wealth per person), and T technology (that is, impact on the environment per wealth per person). These three variables were carefully defined so that, when multiplied, P, A, and T would cancel out of the equation, leading to the identity $I = I$; conversely, P, A, and T may themselves be decomposed to highlight additional subfactors as long as all factors may still cancel out to yield $I = I$. Another useful characteristic of the I=PAT equation stemmed from the fact that the relationship among P, A, and T was multiplicative; consequently, a rise or fall in any of these variables had the potential to change I. For I to rise, PAT would have to increase, and, for I to decline, PAT would have to decrease. For example, if, during one year, population rose by 2% and affluence rose by 4%, T would have to decrease by 6%–in other words, the impact on the environment per unit of wealth per person would have to decline–simply to prevent I from growing.

Until recently, human population had grown robustly for hundreds of years. Worldwide population grew by about 2% in 1970[22] (the date the I=PAT equation was conceived), while, in the same period, the world economy (a measure of affluence) grew by about 8% per capita. [22,23] Though complicated to measure, it is unlikely that the annual rate of technological improvement, as measured by the T factor, has ever approached 10% since the Industrial Revolution. Simply put, the P and A factors have routinely outrun improvements in technology that reduced impact. Nevertheless, since 1970, growth in both population and affluence have slowed substantially. These changes, combined with the invention of powerful new technologies discussed below, offer the possibility that T may soon more than counterbalance the effects of P and A, driving I in a desirable direction: downwards.

How might this occur? The technological improvements discussed later in this article–most prominently, clean energy technologies and artificial intelligence–offer a pathway to dramatic reductions in carbon emissions and in other environmental harms, and the possibility of staggering transformations across a wide swath of human activity. These changes may be sufficient to cause humanity to enter an age in which the T factor in I=PAT grows at a rate large enough to withstand (continued but slowing) growth in both P and A.

## The Dual Nature of Technology

Technology is usually created to fix some sort of problem. As anthropologist Joseph Tainter has written: "Over the past 12,000 years, we have responded to challenges with strategies that cost more labor, time, money, and energy, and that go against our aversion to such costs. We have done this because most of the time complexity works. It is a basic problem-solving tool. *Confronted with problems, we often respond by developing more complex technologies*, establishing new institutions, adding more specialists or bureaucratic levels to an institution, increasing organization or regulation, or gathering and processing more information."[24] (Italics added.) These technologies have taken various forms, from methods of controlling fire, to novel hunting implements, to wheeled vehicles, to subsistence and industrial agriculture. These innovations have provided abundant benefits to humanity, as evidenced by our rapid population growth over the past several thousand years,[22] and by our spread to nearly all areas of the Earth.

The benefits of contemporary technologies can be seen across many facets of life in industrial civilizations, from longer lives,[25] to greater information access,[26] to enhanced mobility.[27] These technological systems have led to standards of living higher than at any other time in human history.[28]

And, as new technologies become available, they often substitute for existing ways of living (see Fig. 1, showing the substitution of cars for horses in the US in the 1900s).

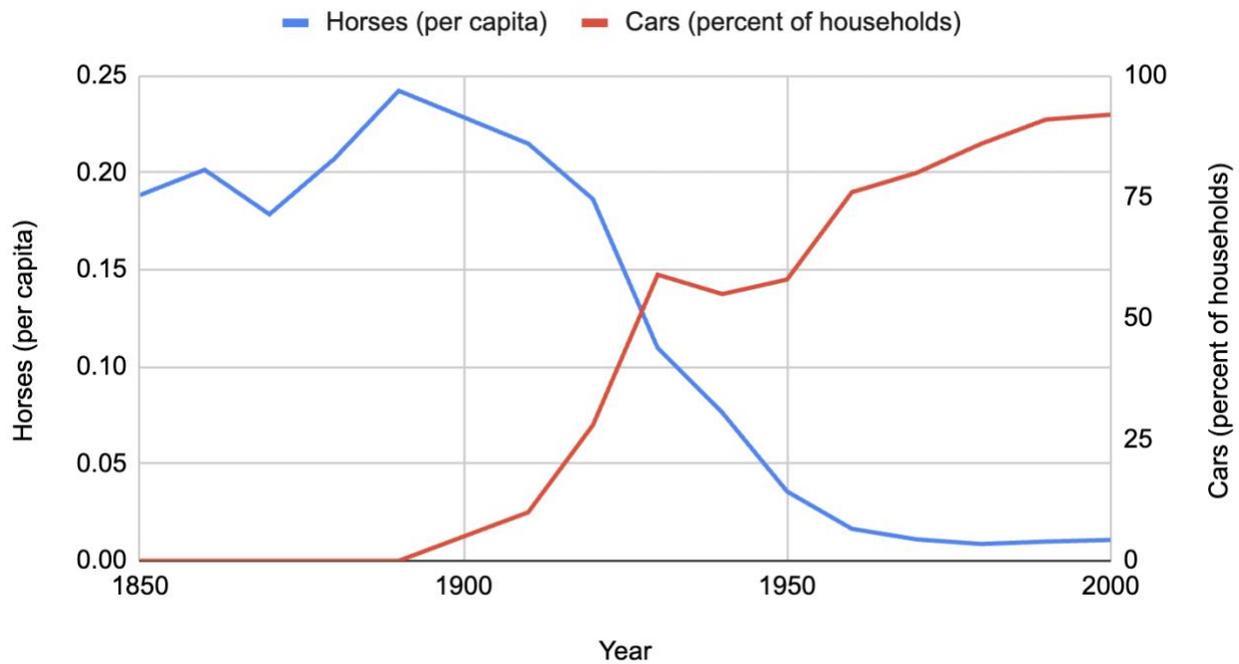

**Figure 1: This chart shows the substitution of cars for horses that occurred in the US in the 1900s. Many technologies of the 1800s---carriages, carts, plows, etc.---were often horse powered, and thus the number of horses serves as a proxy for the prevalence of such technologies.** [29–32]

Technologies are often framed as force multipliers for human activities (e.g., in military,[33] healthcare,[34] and many other domains). The sustainability domain is no exception, with technological innovations working alongside existing human efforts to preserve the environment and improve long-term wellbeing.[15]

Particular technologies, however, have been and continue to be accountable for creating profound societal costs as well. Technological innovations have supported human efforts at hunting and agriculture for thousands of years, at the cost of the overexploitation of many animal species[35] and deforestation[36] around the world. Digital technologies have enabled human communication and connection, but have perpetuated biases that are harmful to millions of people.[37] An array of technologies have contributed to pervasive health risks for billions of people[38] (even though the broader arc of technology appears to bend toward greater health, as evidenced by the trend toward increased human lifespan[25]). And contemporary technologies, *en masse*, are contributing greatly to climate change, biodiversity loss, and many other environmental issues.[39]

The potential for harm has sometimes served to curve innovation in human societies. For example, in Medieval Europe, numerous laws prohibited specific types of innovation. For example, in 14th Century France, "[t]o ensure that no one gained an advantage over anyone else, commercial law prohibited innovation in tools or techniques…"[40] Similarly, a social movement, named the "Luddites", arose during the Industrial Revolution in the United Kingdom.[41] Though the aims of the movement were somewhat complex, the Luddites challenged innovations, such as the Spinning Jenny, that allowed one person to do the work of several. The English romantic poet Lord Byron himself made passionate speeches in the British Parliament, hoping to convince the government to ban such technological advances as harmful to laborers.[42]

While the benefits of technology often serve those creating the technology or funding its development, the costs are often accrued by people other than the creators, by non-human species, or via diffuse effects across long periods of time and/or space. These costs are sometimes referred to as "negative externalities" by economists. For example, the invention of the internal combustion engine allowed humans to travel and transport products rapidly over long distances, but drove demand for fossil fuels whose mining caused habitat destruction, and whose combustion emitted damaging pollutants like sulfur dioxide, nitrogen oxides, lead, particulate matter, carbon monoxide, and carbon dioxide.[43] The acronym NIMBY, meaning "Not In My Back Yard", reflects a sentiment that many people prefer not to have infrastructures (e.g., roads, train tracks) or harmful byproducts (e.g. nuclear waste) located or stored in areas near where they live.[44] The concept of "sacrifice zones,"[45] too, reflects the phenomenon of people in power sacrificing the wellbeing of regions and communities distal to themselves by causing or allowing pollution and other harmful materials to accumulate there.

Nevertheless, many current human cultures embrace, and are characterized by, technological innovation. For example, the vast majority of countries in the world are members of technology-promoting treaties such as the Paris Convention on the Protection of Industrial Property, the Patent Cooperation Treaty, and the World Trade Organization Trade-Related Aspects of Intellectual Property agreement.[46,47]

Technological innovations are often energy intensive.[24] Modern technology has a particularly close relationship with one particular energy source–fossil fuel–that is both vastly powerful and vastly impactful in terms of pollutants such as carbon dioxide.[48] Looking backwards in time, technologies powered by fossil fuels, such as the steam engine, enabled the industrial revolution and its attendant vast increase in human living standards, but also facilitated social stratification through wealth accumulation and health effects, such as respiratory illnesses, across all social strata.[49] Currently, internal combustion engines enable mass mobility and global supply chains, through cars, trucks, ships, and trains; however, they are heavily implicated in a dramatic rise in CO2 concentrations in the atmosphere and accompanying climate change.[39] Many other fossil-fuel-powered technologies have spread rapidly through the industrialized world over the last century as well (see Fig. 2).[29] Climate change is one of the most urgent and important environmental threats currently facing Earth. Climate change is causing sea level rise, threatening to displace hundreds of millions[50] or even billions[51] of people, and altering myriad habitats for organisms worldwide. Climate change is impacting growing seasons and precipitation patterns,[52] which could undermine major elements of humanity's food infrastructure and food security.[53] And climate change is creating extreme weather events, which create profound harm for both humans and non-humans.[54,55] Carbon emissions are also a major factor in ocean acidification.[56] While there are important efforts afoot to decarbonize many aspects of human civilizations,[9,10] carbon emissions from human technologies and activities are still rising,[57] and the effects of such emissions are likely to persist for many decades after any emissions slow-down, or even outright decrease, that may occur.[58]

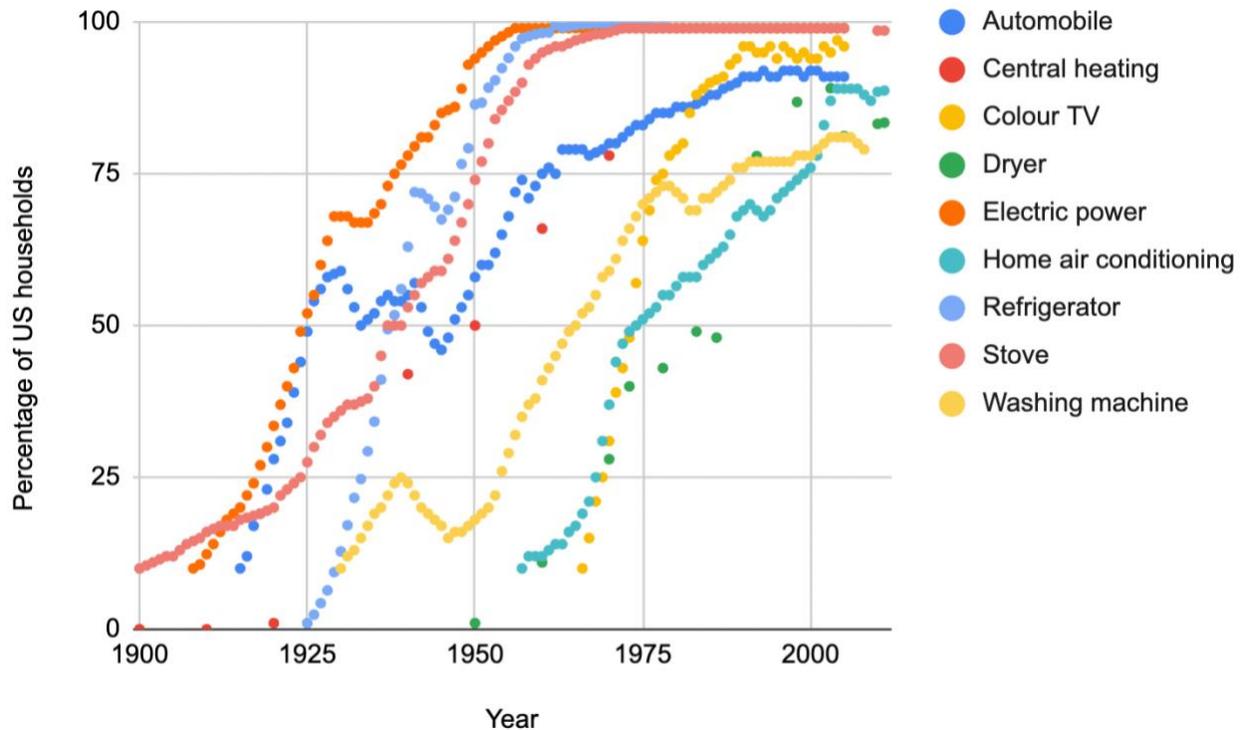

**Figure 2: In the past century, fossil-fuel-intensive technologies have become ubiquitous in the US.[29]**

Fossil-fuel-powered technologies have also enabled the large-scale conversion of the Earth's ecosystems for use for agriculture and other human purposes. But many of these transformations have come at the cost of the forced migration of indigenous populations,[59] the disruption of non-human populations that resided there long before human occupation,[60] and the compromising or complete destruction of existing ecosystem services.[61] This conversion of ecosystems is a key factor in biodiversity loss and deforestation.[36] Many thousands of species are being threatened by anthropogenic environmental effects, potentially leading to what some scientists are calling the Sixth Great Extinction.[62] Biodiversity loss is also implicated in the spread of diseases such as Ebola.[63] In fact, biodiversity loss and climate change appear to affect one another substantially. Humans are transforming the Earth, incidentally creating great harm to the other species with whom we share the planet, to other humans in the future, and in many cases to other humans currently alive.

# Technology Is Implicated in Environmental Harm

While a previous Scientists' Warning article has placed affluence in the key role causing such environmental transformations,[4] technology is nevertheless heavily implicated in these issues. Most of the harm arises from externalities of technologies; for example, fossil fuel companies do not seek to cause climate change via their emission of greenhouse gasses, but have traditionally not borne the costs of that harm due to unwillingness to sacrifice profits.[64] However, some forms of environmental harm are the express purpose of particular social or technological interventions; for example, the eradication of "pests" and "vermin" is typically very much intentional, as evidenced by the Australian policy that contributed greatly to the extinction of the thylacine.[65]

Technological innovators may view themselves as neutral participants in the economy, but they are nevertheless complicit in the harms they propagate. We need to transform our cultures, politics, and societies to address these issues. And we need to transform how technologies amplify the impacts of our actions. In the remainder of this article, we use this conceptual basis to look forward toward technological approaches, and new forms of technology, that may help address these issues, and the effects and externalities that they will entail.

# Toward Technological Change

An array of approaches to technology and technological change that have been discussed, many of which have relevance to the transition to a sustainable future. Table 1 provides a framework for understanding these approaches, structured around whether they are primarily concerned with present or future technological systems, and whether they focus on the benefits or costs of those systems. In the following subsections, we group these approaches into several high-level categories, that seek to a) reduce current harms, b) enable future benefits, c) reduce future harms, and d) enable new narratives that point toward sustainable futures.

|  | **Focus on Costs** | **Focus on Benefits** |
|---|---|---|
| **Future technology** | Substitution[66]<br>Value Sensitive Design[67]<br>Technology within Limits[68]<br>Benign technology[69] | Substitution[66]<br>Value Sensitive Design[67]<br>Technology within Limits[68]<br>Technological Evangelism[70] |
| **Present technology** | Substitution[66]<br>Undesign[16] | Substitution[66]<br>Resistance to Change[71] |

**Table 1: A matrix showing various approaches to technological change, grouped based on whether they focus on present technology vs. future technology, and whether they focus on the costs or benefits of those technologies.**

## Reducing Current Harms

A critical goal in addressing climate change is to reduce the use of currently harmful technologies, such as gasoline-powered internal combustion engines and coal-powered electricity plants. While the harm from these technologies are typically diffuse and pervasive rather than acute and local, there is nevertheless a strong scientific consensus that these technologies are heavily implicated in profound, long-term harm via pollution and global climate change.[39] The need for these changes are well established; we refer the reader to energy innovator Saul Griffith's excellent summary of the need to set aside these existing technologies.[9] We would also like to point to existing work on enabling transition pathways to allow human societies to abandon old technologies that are known to be harmful.[16,72,73]

How exactly to disengage with the reduction in existing, widely-used technologies is a topic of some debate. Some activists propose that protests and divestment from fossil fuels are the best path forward.[74]

"Fossil fuel divestment stigmatizes the fossil fuel industry for its culpability in the climate crisis and frames climate change as a moral crisis."[75] Others, such as Saul Griffith mentioned above, take a more conciliatory view: "Climate activists can fight the fossil fuel companies until the end of our lives, or Americans can come together, thank these companies for a century of service, and engage with them in the fight for our future." Regardless of the exact approach, the reduction, mitigation, or even elimination of technological systems with vast harmful externalities is critical to the future thriving of humanity and other species.

We see two main pathways for reducing, mitigating, or eliminating fossil-fuel-based technology. The first pathway involves technological substitution. The replacing of one technological system by another has been a common method for addressing the shortcomings of one technology when another is available.[66] For example, there are promising efforts afoot to substitute fossil-fuel-based systems with electrification and other clean energy technologies.[9] We discuss the benefits and costs of these technologies in a later subsection, and note that much of the electricity used in "clean" technologies, such as electric cars, still currently originates from burning fossil fuels.

The second pathway involves undesign. Undesign involves the "intentional negation of technology,"[16] and "articulating the value of absence,"[72] that is using the tools of design to offer alternative courses of action that do not entail continued use of a technology. Undesign seeks to reduce the usage of a particular type of technology, and as such, could be a useful approach for reducing fossil-fuel-based technologies in some contexts. Nevertheless, undesign is difficult in contexts where people have come to rely strongly on particular goods and services; in such contexts, people may need to be guided to new cultural narratives in which those goods and services are less central. We discuss the need for new narratives below as well.

Both substitution and undesign need to push against resistance to change, which tends to focus on the benefits of the current systems (e.g., "the fact that we value the groups to which we belong, and therefore changing our attitudes or behavior is tantamount to leaving the comfortable embrace of a social reality of which we are a part"[71]), regardless of their harms (especially their long-term, diffuse, or physically or temporally distal harms).

Whether via substitution or undesign, research into how to disconnect, or diminish the impact of, many different processes in industrial civilizations from their underlying fossil fuel technological infrastructures is of paramount importance.

## Enabling Future Benefits

To enable the reduction or phasing out of fossil fuels, it is also critical to continue researching, developing, deploying, and evaluating novel forms of technology that may take its place. The adoption of new technologies may usher in an array of benefits. We use two main classes of technology as elucidating instances of the benefits that may arise from the deployment of new technologies: clean energy technologies and artificial intelligence.

Clean energy technologies include electrification technologies–from solar panels to electric transportation systems to heat pumps in every home. While electrification may not be able to address every aspect of decarbonization (e.g., long-distance aviation is difficult to achieve without energy-dense liquid fuels[10]), large-scale electrification is a key component of "net zero" emissions energy systems.[10] Large-scale electrification will entail substantial changes to industrial infrastructures such as the energy grid; however, it should not require severe austerity or profound alterations in most people's lived experiences.[9] These technologies can allow humans in industrial civilizations to maintain similar standards of living as they do with fossil-fuel powered systems, but less expensively in the long run, and with far lower environmental impacts.[9]

The second class of technology we discuss here is AI. AI does not impinge on near-term carbon emissions as dramatically and immediately as clean energy does, but it also shows great promise across longer time horizons. AI is already being used to enable land use reform via the planning of wildlife corridors,[76] to monitor methane emissions,[77] and to optimize supply chains.[78] Beyond those domains, AI has the potential to guide humanity to new ways of living that are currently beyond our abilities to conceive, develop, and deploy. AI is currently far less energy-intensive than humans at tasks such as writing and illustration that were previously almost exclusively the domain of human creativity,[79] which could have far-reaching implications for future visions of human civilizations. Similarly, AI is becoming quite competent at writing computer code (see Fig. 3), which could help manage the complexities of 8 billion plus people cohabiting on Earth. While AI is likely to make many human jobs obsolete, it may also enable the coordination of human systems far more effectively than humans have traditionally done.[80] (It is also likely to create entirely new classes of jobs.[81]) It may open entirely new ways of undertaking other tasks, potentially transforming domains from individual wellbeing[82] to international diplomacy.[83] While current AI systems are not without their challenges (e.g., they are often wrong[84]), there are efforts afoot to connect different types of AI systems together (e.g., large language models with knowledge graphs[85]) to overcome the limitations of each genre.

Two technological approaches are relevant here: a "technology within limits" perspective,[68] and the value sensitive design (VSD)[67] framework. While the "limits" perspective was developed within the computing field, we see it as being relevant to technology more broadly. We present here a modified version of two of the key insights presented by Nardi et al. in their foundational article on the topic:

> We [propose] that [technology fields] transition toward '[technology] within limits,' exploring ways that new forms of [technology] may support well-being for both humans and non-human species while enabling human civilizations to live within global ecological and material limits. [Technology] underlies virtually all the infrastructure of global society, and will therefore be critical in shaping a society that meaningfully adapts to global limits. (Adapted from Nardi et al.[68])

As technology seeks to serve such long-term goals, we encourage technological innovators to engage with their work using a "limits" perspective. (A limits perspective may be juxtaposed with "technological evangelism,"[70] a term often used for an approach primarily focused on the benefits of technology; while evangelism is sometimes seen as being uninterrogated and "hype" focused, it may be a useful approach as well, for example, in overcoming resistance to change.)

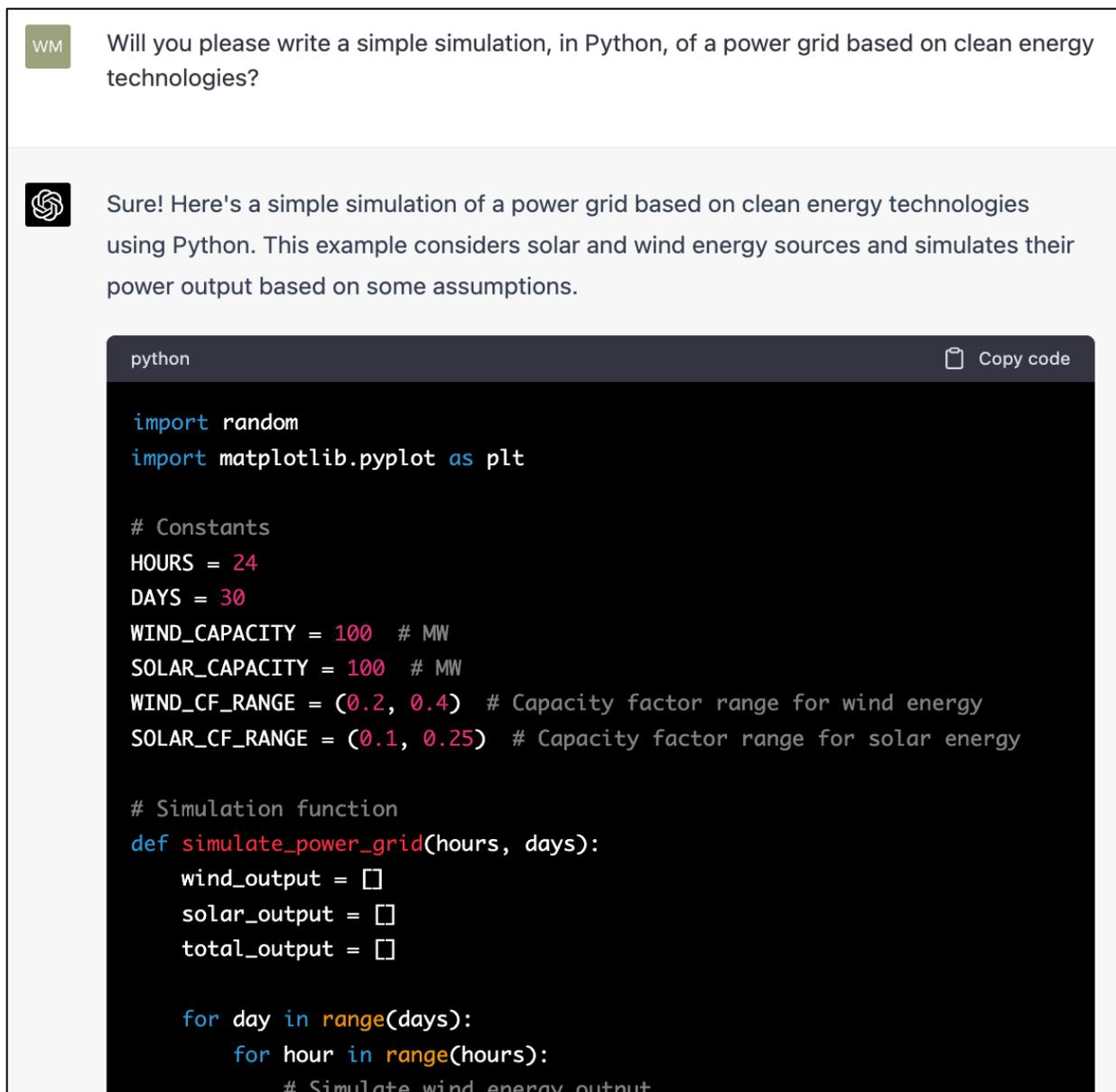

**Figure 3: This figure shows output from the ChatGPT AI system running GPT-4 creating a simple simulation of a clean energy-based power grid.**[86]

VSD[67] offers a framework for thinking about the benefits (and also the harms) of future technological systems. VSD integrates with the substitution approach discussed earlier, as VSD seeks to foster the development of technological changes that are aligned with human values, both in terms of the new systems that will be adopted, and the old systems that will be supplanted.

Despite the great potential for these technologies across many domains, we nevertheless caution against unbridled techno-utopianism. Technological development is famous for failing to deliver on "quick fixes" and other promises of change.[14] While the pressing demands of environmental crises encourage a vigorous pursuit of technological change, we now want to turn to the challenges implicated in these potential future technologies, and encourage vigilance against likely future problems that they may pose.

**Reducing Future Harms**

While broad adoption of clean energy technologies has powerful benefits, given the need to address climate change, such technologies are likely to create or exacerbate some environmental and social problems in the future, such as human rights abuses associated with cobalt mining,[87] proliferation of electronic waste,[88] and short-term job displacement (although it is likely to create even more jobs on a longer time horizon).[9] There are almost certainly other challenges, decades in the future, that will not become apparent until electrification has become as pervasive as fossil fuel-based energy systems now are. While it is critical in the present to electrify as rapidly as possible to maintain high standards of living while phasing out fossil fuels, it nevertheless remains relevant with electrification, as with all human technologies, to remain vigilant about future harms that may arise from this new energy system.

The long-term risks of the proliferation of AI are perhaps more broadly concerning. One possible future harm involves the runaway use of energy. Regardless of how much clean energy we may be able to derive from broad-scale electrification, future AI may be able to use it in the quest for ever more highly optimized human processes, ever finer personalization of content, and other human goals. Current AI models require energy on par with several car lifespans to train them;[13,89] regardless of how large future energy resources may be, future AIs will almost certainly be able to increase its energy uses until it is operating in a context of limits.[68] (We note that, at present, the carbon emissions of AI engaging in many tasks is still far less than the emissions produced by humans doing the equivalent work.[79])

A second concern involves the problem of autonomous AIs becoming independent and powerful enough that they pose a real danger to humanity. This concern is sometimes explored under the moniker of the "alignment problem".[90,91] The alignment problem involves understanding how to align the goals of a very powerful AI with the goals of its creators, or of humanity more broadly.[90,91] Science fiction has been rife with instances of dangerous AI superintelligences, from The Terminator to The Matrix; nevertheless, concern with the alignment problem is now entering more traditional scientific discourse as AI becomes more powerful.[90,91]

In addition to these existential concerns, AI could propagate an array of other harms, from perpetuating biases,[37] to overoptimizing systems in violation of human values,[92] to empowering crime.[93]

While we identify here an array of potential future harms from both electrification and AI, we believe that these harms are substantially less problematic than these technologies' potential benefits from addressing the pressing environmental crises that humanity is currently facing.

To address the externalities that may arise from both of these forms of technology, technologists must engage with design processes that grapple with those issues from the outset, rather than seeing them as a secondary concern. Design processes that take many stakeholders into account, including nonhuman species and ecosystems,[94] have a greater chance of doing so effectively.[95]

One approach to addressing the harms of future technologies is a "benign technology" perspective.[69] This perspective also arose out of computing; we present an adapted version here, expanded to encompass all technology:

> We propose one possibility: benign [technology], a general design framework for building [technological] systems that are less likely to produce harmful impacts to the ecosystem (and thus to human society) and are less likely to become trapped by Sevareid's Law (that "the chief source of problems is solutions"). (Adapted from Raghavan.[69])

The benign technology principle involves exerting effort to identify and address externalities. "[W]hen we do build things, we should engage in a critical, reflective dialog about how and why these things are built."[72] VSD[67], discussed earlier, also addresses the prospect that preventing harm from future technologies should be central to the process of technological innovation.

We encourage technologists to engage with both the benefits and the harms of the technologies they are creating as well as of those they are supplanting.

**Enabling New Narratives**

Finally, researchers should seek to create and study technologies that teach and inspire people to consider new narratives that could underpin how our civilizations interact with the ecosystems in which they are embedded. While some efforts at enacting change may allow for human lives in the industrialized world to continue largely unchanged,[9] there are also important calls for new cultural narratives that do not involve the many forms of environmental harm that are closely intertwined with modern market economies. As environmental journalist George Monbiot has written, "[w]e have to come together to tell a new, kinder story explaining who we are, and how we should live."[96] Technology can help articulate such new stories, and teach them to potentially billions of people.

The question of what narratives are needed relates to the two exemplar technology domains we have discussed above. Clean energy technologies are likely to integrate with existing lifestyles in the industrialized world sufficiently that they may not entail substantial shifts in how we live our lives. However, to realize the potential of AI, we may need to undertake substantive shifts in lifestyle. We will need to develop new cultural narratives that align with sustainable futures.[97] Hopefully, these shifts will be in the direction of higher quality of life. "The only way you can change a story is to offer a new one. And you can do so only by producing a better story."[97]

An approach called design fiction (defined as "the deliberate use of diegetic prototypes to suspend disbelief about change"[98]) can be used to help envision new futures. Design fiction has been used to envision sustainable futures in particular.[99] This approach to the design of technology can help people think in a "different conceptual space"[98], and thereby get past the confines of present technologies and present cultural norms and expectations to conceptualize the transition to sustainability.

We issue a global appeal for technology developers and researchers to develop new technologies, and for artists, designers, and storytellers to develop new narratives, that help civilizations integrate these technologies in ways that improve human lives, the lives of non-human species, and the prospects for the future of life on Earth. Many people feel that technology will save us. But it will only save us if we develop cultural norms that allow it to do so. "[O]nly widespread changes in norms can give humanity a chance of attaining a sustainable and reasonably conflict-free society."[100]

## Scientists' Warnings

This article is a warning, but it is a warning in three parts. The first part warns that current technologies are causing profound environmental and social harm. It is critical to phase out these technologies as soon as possible.

The second part warns that, as human civilizations deploy new technologies in the process of phasing out current technologies, it is important to remain vigilant for potential future harm, and attempt to reduce that harm as much as possible. Nevertheless, despite their potential to cause various forms of harm, we strongly believe that the potential for future technologies, and in particular clean energy and AI technologies, to enable human civilizations to address the current, dire environmental problems such as climate change and biodiversity loss are well worth the future technological risks these systems may engender. We offer this warning because there is a real risk that human civilizations will fail to transform their energy systems, infrastructures, cultures, politics, and societies fast enough to avert the potentially catastrophic impacts of environmental disruption and societal collapse. Therefore, this warning is here to warn humanity not to fear these changes, and not to fail to make these changes, but instead to embrace the possible futures that these technologies may enable.

The final part of the warning is that, while technology is very powerful, and has been instrumental in shaping many aspects of the world humans have made over the past several thousand years, technology alone will likely be insufficient to enable a transition to a sustainable future. Sustainable futures will require profound shifts in culture, politics, and society more broadly.[15] We hope that the myriad technological fields will support these transitions as effectively as possible, rather than becoming entrenched in business as usual. Sustainable futures can hopefully support the long-term wellbeing of humanity and many other species, and technology can hopefully help bring these futures into existence.

We propose that the following is an admirable goal for this effort: the indefinite continuation and ongoing well-being of the human species and other currently existing species. This goal is unachievable, since various individuals and species are inherently in conflict, via food webs, competition for habitat, etc. Nevertheless, we see it as a target at which to aim our efforts. And we anticipate that technological change will be integral to working toward that goal.

## Author Contributions

B.T. and A.T. conceived the project. B.T. led the writing of the manuscript and creation of the figures. A.T. wrote the I=PAT section and contributed content and revisions throughout the manuscript and figures. W.R. contributed content and revisions throughout the manuscript and figures. All authors reviewed and confirmed the final manuscript.

## Acknowledgments

This material is based upon work supported by the US National Science Foundation under Grant No. DUE-2121572.

## Competing Interests

The authors declare no competing interests.